\documentclass{article}

\usepackage{arxiv}

\usepackage[utf8]{inputenc} 
\usepackage[T1]{fontenc}    
\usepackage{hyperref}       
\usepackage{url}            
\usepackage{booktabs}       
\usepackage{amsfonts}       
\usepackage{nicefrac}       
\usepackage{microtype}      
\usepackage{lipsum}		
\usepackage{graphicx}
\usepackage{natbib}
\usepackage{doi}
\usepackage{threeparttable}

\title{EEGLog: Lifelogging EEG Data When You Listen to Music}

\author{{\includegraphics[scale=0.06]{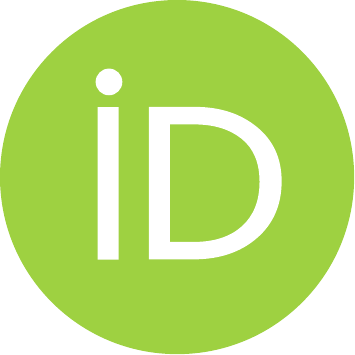}\hspace{1mm}Jiyang Li} \\
	Department of Computer Science and Engineering\\
	University at Buffalo, State University of New York\\
	Buffalo, NY, 14260, USA \\
	\texttt{jiyangli@buffalo.edu} \\
	\And
	{\includegraphics[scale=0.06]{orcid.pdf}\hspace{1mm}Ann Gina Konnayil} \\
	Department of Computer Science and Engineering\\
	University at Buffalo, State University of New York\\
	Buffalo, NY, 14260, USA \\
	\texttt{annkonna@buffalo.edu} \\
    \And
	{\includegraphics[scale=0.06]{orcid.pdf}\hspace{1mm}Adam Russell} \\
	Department of Computer Science and Engineering\\
	University at Buffalo, State University of New York\\
	Buffalo, NY, 14260, USA \\
	\texttt{adrussel@buffalo.edu} \\
    \And
	{\includegraphics[scale=0.06]{orcid.pdf}\hspace{1mm}Dingran Wang} \\
	Motorola Solutions Inc.\\
	Schaumburg, IL, USA\\
	\texttt{dingran.wang@motorolasolutions.com} \\
    \And
	{\includegraphics[scale=0.06]{orcid.pdf}\hspace{1mm}Yincheng Jin} \\
	Department of Computer Science and Engineering\\
	University at Buffalo, State University of New York\\
	Buffalo, NY, 14260, USA \\
	\texttt{yincheng@buffalo.edu} \\
    \And
	{\includegraphics[scale=0.06]{orcid.pdf}\hspace{1mm}Seokmin Choi} \\
	Department of Computer Science and Engineering\\
	University at Buffalo, State University of New York\\
	Buffalo, NY, 14260, USA \\
	\texttt{seokminc@buffalo.edu} \\
    \And
	{\includegraphics[scale=0.06]{orcid.pdf}\hspace{1mm}Zhanpeng Jin} \\
	Department of Computer Science and Engineering\\
	University at Buffalo, State University of New York\\
	Buffalo, NY, 14260, USA \\
	\texttt{zjin@buffalo.edu} \\
}

\date{}

\renewcommand{\shorttitle}{EEGLog: Lifelogging EEG Data When You Listen to Music}

\hypersetup{
pdftitle={A template for the arxiv style},
pdfsubject={q-bio.NC, q-bio.QM},
pdfauthor={David S.~Hippocampus, Elias D.~Striatum},
pdfkeywords={First keyword, Second keyword, More},
}

\begin{document}
\maketitle

\begin{abstract}
Self-tracking has been long discussed, which can monitor daily activities and help users to recall previous experiences. Such data-capturing technique is no longer limited to photos, text messages, or personal diaries in recent years. With the development of wearable EEG devices, we introduce a novel modality of logging EEG data while listening to music, and bring up the idea of the neural-centric way of life with the designed data analysis application named EEGLog. Four consumer-grade wearable EEG devices are explored by collecting EEG data from 24 participants. Three modules are introduced in EEGLog, including the summary module of EEG data, emotion reports, music listening activities, and memorial moments, the emotion detection module, and the music recommendation module. Feedback from interviews about using EEG devices and EEGLog were obtained and analyzed for future EEG logging development. 
\end{abstract}


\section{Introduction}

\begin{figure*}
  \includegraphics[width=\textwidth]{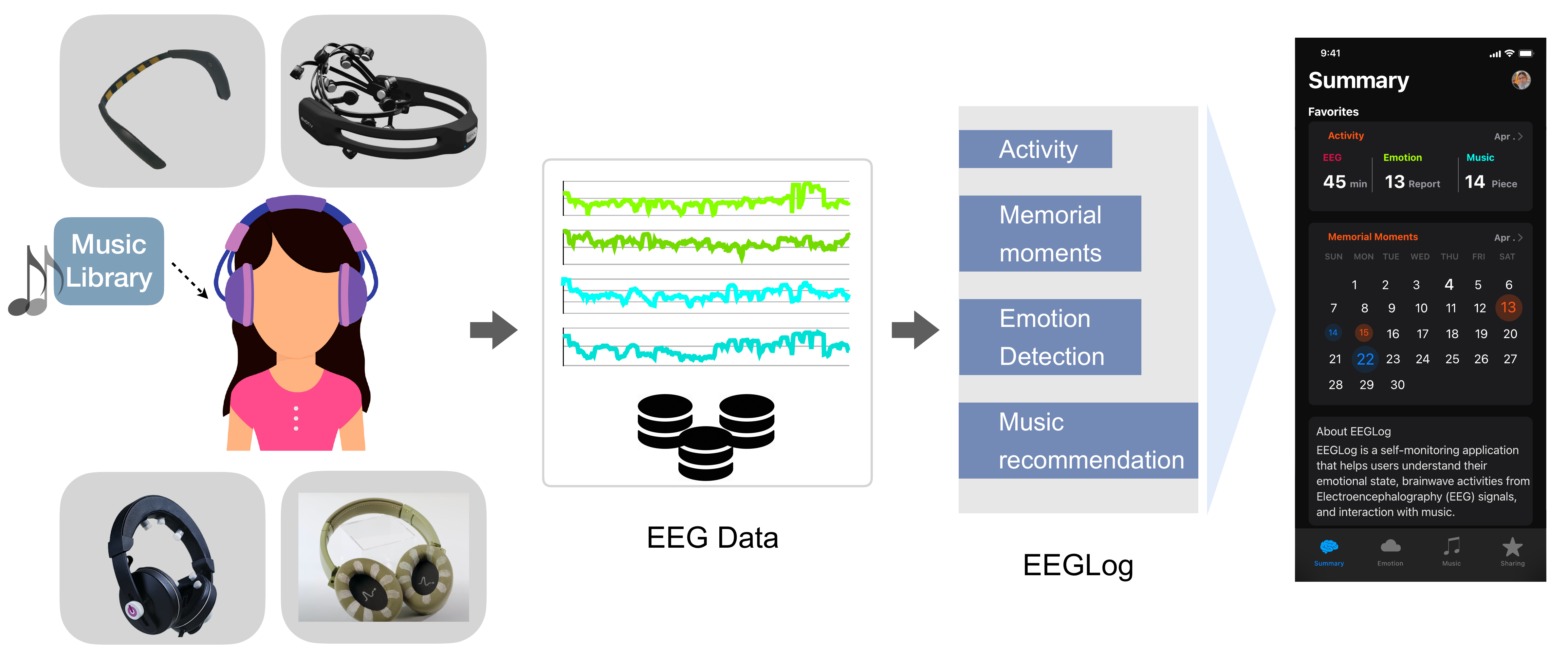}
  \caption{Overview of EEGLog. Users use an EEG device to log in their EEG data while listening to music and report their emotional states afterward. The application named EEGLog is designed to show past data, detect emotion, and recommend music for users' self-monitoring and reflection.}
  \label{fig:over}
\end{figure*}

Lifelogging and ``digital memories'' (or ``e-memory'') have been introduced by researchers ten years ago by capturing users' daily information like emails, website histories, images, and so on \cite{bell2009total, sellen2010beyond}. It compensates for the fallible human memory and enables us to use, recall, and share our old experiences. With the advancements of various consumer-grade wearable sensors and devices, users' physiological and behavioral data can also be tracked for health and emotion monitoring \cite{mcduff2012affectaura}. Among them, the emerging, consumer-grade user-wearable EEG device that is introduced in recent years recording people's brain activity and revealing the mechanisms of human memory and emotion, has great potential to be logged in human's daily life in Human-Computer Interaction (HCI) researches.

Self-tracking technologies provide methods of improving reflective thinking and self-awareness \cite{li2011understanding, rivera2012applying}. And Sanches et al. \cite{sanches2019hci} indicated that systems developing self-tracking can benefit people with affective disorder and strengthen the communications between clinicians and patients. Emerging wearable devices and data analysis techniques helps people self-monitoring that can reflect their behaviors. An example of such a trend is called the Quantified Self (QS) movement \cite{lee2013quantified}, which has followers eager to increase their understanding of themselves by self-tracking. People nowadays are affected by anxiety, depression, exhaustion, and other emotional problems. The self-monitoring and reflection can provide proper emotional cares for people with needs. Technologies for self-reflection and emotional well-being applications are attracting more attention. Meanings of lifelogging and self-tracking are summarized by Sellen and Thittaker \cite{sellen2010beyond} into five Rs: Recollecting, Reminiscing, Retrieving, Reflecting, and Remembering. Those potential values are highly involved in human's daily life but lack explicit descriptions. 

St\r{a}hl et al. \cite{staahl2009experiencing} introduced the Affective Diary, which combined data from body sensors and activities on the mobile phone like text messages and photographs, to provide cues of emotional expressivity and reminiscence of bodily experiences. Isaacs et al. \cite{isaacs2013echoes} built a smartphone application named Echo and showed that the technology-mediated reflection (TMR) could improve well-being. The application recorded users' everyday experiences, including their writing of a subject line, rating of the happiness, photos, videos, etc., and then created reflection posts on smartphones for users to re-rate. McDuff et al. \cite{mcduff2012affectaura} designed a reflective tool called AffectAura to explore the potential reflective power, which was offered by pairing the affective data with the knowledge of workers’ information and data interaction artifacts, and showed the effect that users could reason forward and backward their emotions. Vaara et al. developed Affective Health to help people monitor and understand the stress level from their every day activities by visualizing the biofeedback designed as `spirals' \cite{vaara2010temporal}. To record event instances in time, Chong et al. showed SqueezeDiary that uses squeeze gestures to trigger the denoting behavior \cite{chong2014squeeze}.


Despite the increasing interest and advances in lifelogging techniques, many technical challenges remain and need to be overcome, such as data storage and privacy, dependence on human operations, and data analysis and usage. To address the problem of limited data storage and record the users' valuable moments, it is imperative to activate the lifelogging service on demand. Instead of relying on the users' manual operations and activation of the lifelogging, we propose to use the music listening as a lifelogging activation trigger, which involves both the device (i.e., the consumer wearable EEG device can be worn by the general population) and the personal emotional tendency (i.e., the user's emotional moments associated with music can be recorded). Besides, as the media file other than monotonous digital data, music can serve as more vital cues or reminders for lifelogging, instead of only providing content to be remembered. And music, as being extensively studied in literature \cite{Koelsch2014, Taruffi2017, Siddharth2019, Ehrlich2019}, is proven to be one of the most effective and accessible manners to evoke emotions and influence moods \cite{goldstein1980thrills}.

To this end, we propose a new, convenient and accessible lifelogging scheme that incorporates electroencephalograph (EEG) sensing with the commodity headphones, so that users' brainwaves can be collected while they are listening to music. In such an intuitive way, the personal brainwave library can be built and the data will be used for tracking the users' emotional states and music interactions through the mobile apps we developed. To assess the applicability and effectiveness of the proposed personal music-EEG library approach (EEGLog) in lifelogging, we aim to examine the performance of EEGLog using multiple commercial off-the-shelf (COTS), consumer-grade wearable EEG devices, as shown in Figure \ref{fig:over}, including MUSE 2, Emotiv EPOC+, mBrainTrain Smartfones, and Neurable Enten. We separated them into four independent experiments with the same procedure. After collecting users' EEG data over a considerable period of time, ranging from a few days to even weeks, the collected EEG dataset and corresponding emotion analysis are provided by EEGLog for self-reflection. For emotion detection and regulation algorithms, Hollis et al. \cite{hollis2018being} conducted a survey that raised a phenomenon resulting from the algorithms which would override users' own interpretations even when they are conflicted. Thus we design the EEGLog prototype based on principles of different levels of data manipulation: no manipulation, primitive functions, and machine learning (ML) models. Specifically, we have made the following contributions: 
\begin{itemize}
    \item We proposed a novel emotion-lifelogging approach based on EEG that seamlessly interacted and incorporated with ordinary music listening in people's daily life.
    \item We developed the proof-of-concept EEGLog application and provided users with three modules for self-monitoring and reflection.
    \item We investigated the applicability and effectiveness of the proposed EEGLog approach on four popular COTS wearable EEG devices and summarized users' comments of experience for future research.  
\end{itemize}

\begin{table*}[tbp!]
\centering
\caption{Specifications of Four EEG Devices}
\label{tab:devices}
\begin{threeparttable}
\begin{tabular}{lllll}
\toprule
 & \textbf{Emotiv EPOC+} & \textbf{MUSE 2} & \textbf{mBrainTrain Smartfones} & \textbf{Neurable} \\ \hline
\textbf{Electrodes} & Semi-dry & Dry & Semi-dry & Semi-dry \\ \hline
\textbf{Channels} & 14 & 4 & 12 & 20 \\ \hline
\textbf{Locations} & \begin{tabular}[c]{@{}l@{}}AF3 F7 F3 \\ FC5 T7 P7\\ O1 O2 P8 T8 \\ FC6 F4 F8 AF4\end{tabular} & \begin{tabular}[c]{@{}l@{}}TP9 \\ AF7\\ AF8\\ TP10\end{tabular} & \begin{tabular}[c]{@{}l@{}}L1 - L4\\ R1 - R4\\ C3 C4 Cz\end{tabular} & \begin{tabular}[c]{@{}l@{}}1 - 10\\ 11 - 20\end{tabular} \\ \hline
\textbf{Functionality} & \begin{tabular}[c]{@{}l@{}}Brainware \\ device\end{tabular} & \begin{tabular}[c]{@{}l@{}}Meditation\\ headband\end{tabular} & \begin{tabular}[c]{@{}l@{}}EEG \\ headphones\end{tabular} & \begin{tabular}[c]{@{}l@{}}EEG\\ headphones\end{tabular} \\ \hline
\textbf{Sampling rate} & 128 Hz & 256 Hz & 500 Hz & 500 Hz \\ \hline
\textbf{Cost} & \$799 + & \$289 + & \$9,000 + & \$399 + \\ \hline
\textbf{Supplement} & Saline solution & Alcohol wipes & Saline solution & Alcohol wipes \\ \hline
\begin{tabular}[c]{@{}l@{}}\textbf{SDK}\end{tabular} & EmotivPro & \begin{tabular}[c]{@{}l@{}}Mind\\ Monitor\end{tabular} & \begin{tabular}[c]{@{}l@{}}Smarting \\ Streamer\end{tabular} & \begin{tabular}[c]{@{}l@{}}CGX \\ Acquisition\end{tabular} \\ \hline
\textbf{Setup time\tnote{a}} & $\sim$ 10 mins & $\sim$ 5 mins & $\sim$ 12 mins & $\sim$ 5 mins \\ \hline
\begin{tabular}[c]{@{}l@{}}\textbf{Comfortable} \\ \textbf{wearing time}\tnote{b}\end{tabular} & $\sim$ 28 mins & $\sim$ 36 mins & $\sim$ 20 mins & $\sim$ 39 mins \\ 
\bottomrule
\end{tabular}
\begin{tablenotes}
        \footnotesize
        \item[a] The data is based on the overall experiment setup time from the experimenters' personal experiences.
        \item[b] The data is based on the feedback of human subjects participating in this study.
\end{tablenotes}
\end{threeparttable}
\end{table*}

\section{Related Work}
\subsection{Lifelogging \& Personal Informatics}
Lifelogging has been introduced as a complete record, or total capture of users' life \cite{gemmell2009memory}. And it can enable people to look back over their lives and search through past experiences. A study reported by Sellen et al. \cite{sellen2007life} provides evidence that some kinds of cues captured by life-logging technologies can be shown to provide effective links to events in people‘s personal past. However, exhaustive records have been criticized in the past for reasons that personal digital records are not equally valuable and they can overwhelm maintaining and retrieving meaningful information from large archives \cite{sellen2010beyond}. Alternatively, targeting moments that are meaningful and things users want to remember provides the greatest utility of a lifelogging system. 



St\r{a}hl et al. introduced the Affective Diary, which combined data from body sensors with other types of media, and provided diary entry creation by incorporating abstract representations of mood and activity (cognitive and physical experiences) with data streams from the user’s phone and hand-written entries. The tool was designed for the active daily creation of diary entries, and the authors studied how such multimodal data incorporation involving affective data enhanced the creation experience \cite{staahl2009experiencing}. McDuff et al. introduced the AffectAura to explore the potential reflective power that might be offered by pairing affective data with the knowledge of workers’ information and data interaction artifacts. They specifically designed transparency into the system so that participants could view and reflect upon the relative affective measures predicted by the system, namely, positive vs. negative valence, low vs. high arousal, and low vs. high engagement. \cite{mcduff2012affectaura}. 

\subsection{EEG, Music, \& Emotion}



Electroencephalograph (EEG) has been extensively studied to reveal people's identification, comprehension, awareness, emotion, and so on \cite{armstrong2015brainprint, schneegass2020braincode, takahashi2004remarks, andres2020introducing, lin2010eeg, alzoubi2009classification}. Specific frequency bands of EEG signals can be inferred for different brain activities \cite{dasdemir2017analysis}. Many prior studies have declared the discrimination of valence emotions by neural responses, as positive and negative musical stimuli elicit different cortical lateralization patterns \cite{altenmuller2002hits, schmidt2001frontal, flores2007metabolic}. It has been well studied and recommended that, as a common practice, the power spectra of the EEG can be divided into five bands --- delta ($\delta$: 1-3 Hz), theta ($\theta$: 4-7 Hz), alpha ($\alpha$: 8-13 Hz), beta ($\beta$: 14-30 Hz), and gamma ($\gamma$: 31-50 Hz) \cite{mantini2007electrophysiological}. 

Unfortunately, most of the prior EEG-related studies primarily focused on the investigation of conventional wired EEG caps, which have bucky form factors and restricted application scenarios in the lab. This has significantly limited the wide deployment of EEG-based brain-computer interface (BCI) solutions. In the past decade, there have been increasing attention and interest in truly affordable, accessible, and usable EEG devices that be conveniently used in people's daily life, and thus a variety of such wearable EEG devices have become available in the market \cite{electronics9122092,10.3389/fninf.2020.553352}. In addition to the traditional EEG electrode placements (e.g., forehead or full-head), around-the-ear EEG collection has also been explored recently, which extends the coverage of the head for a more practical usage in daily life \cite{bleichner2017concealed, debener2015unobtrusive, alcaide2021eeg}, and brings mobile EEG to a new stage of transparent EEG that works as an auxiliary function of headphones rather than a single device. It is proven to be sensitive to picking up electrical signals like P300 and alpha oscillations, compared with scalp EEG signals. Also, sensors are placed on the hair-free skin around the mastoid bone and are partly located over the inferior temporal cortex, which relates to emotion appraisal and processing \cite{sander2005systems}. 


Music can elicit various emotional experiences \cite{hays2005meaning, laukka2007uses}, and it has been shown that certain brain regions are activated in neuroimaging studies \cite{blood2001intensely}. People regulate emotion routinely through music and spend billions annually on music \cite{rickard2004intense, laing2009world}. Studies have been done to recognize music-induced emotions by EEG signals and the strategies people use. Ramirez et al. introduced a musical neuro-feedback approach based on Schmidt's study to recognize the emotions of the elderly \cite{ramirez2015musical}, aiming to increase their arousal and valence based on music selections. The same approach was also applied to palliative care of cancer patients \cite{ramirez2018eeg}. Earlier, Sourina et al. proposed an adaptive music therapy algorithm based on detected emotions, and obtained satisfaction from human subjects towards the therapy music \cite{sourina2012real}. They designed the recommender for six discrete emotions in Russell's v/a model, and participants can assign those emotions to the system to obtain the corresponding music pieces.

\section{Implementation}

\subsection{System Overview}
The recent shift from dedicated research tools to commercially available consumer gadgets makes non-invasive EEG devices more accessible to people in understanding their brain activities, and makes it possible to reflect and interact with their psychophysiological responses. To introduce an EEG-based self-monitoring lifestyle and make EEGLog widely accessible to the general public, we select and test four commercially available EEG devices with the usage of EEGLog. Selected COTS wearable EEG devices were shown in Table \ref{tab:devices}. They represent the most popular and easy-to-wear EEG devices in the market, and two of them are headphone prototypes with embedded EEG sensors. We recruited 24 human participants and collected EEG data over a considerable period (ranging from a few days to even weeks). The same experimental procedure was repeated for four devices independently. EEGLog was designed as a mobile application to show the recorded and analyzed data, including data collection activities, emotion detection, and music recommendations. Modules and interfaces of EEGLog are shown in Figure \ref{fig:over}. 


\subsection{EEG Devices}
Low-cost consumer-grade EEG devices have been growing explosively in recent years. We selected four COTS EEG devices with at least four electrodes. Two of them are traditional scalp EEG devices, and the other two are developed with back ear EEG techniques. Detailed introductions of devices and reasons we chose the device are in the following paragraphs. Other specifications are listed in Table \ref{tab:devices}. Cost with a 'plus' sign refers to the extra cost of software, application subscriptions, and supplements. The high cost of mBrainTrain's Smartfones is because of the fact that the device is still in the prototype stage and hasn't been in large-scale production. However, given its unique headphone-style design and influences in back ear EEG researches \cite{debener2015unobtrusive, bleichner2016identifying, mirkovic2016target, lee2020decoding}, it's worth including this device in the discussion and study of logging EEG data. The average device setup time is based on our operational records in the experiments; the comfortable wearing time is obtained based on the feedback of all participants (i.e., from the time when the user put on the EEG device until when the user started to complain about the uncomfortable sensations caused by the sticky attachment, prickling pain, or tightness of the headset). If a device was not complained about by participants during experiments, one participant took the tolerance test and we recorded the time when the participant reported discomfort.

Emotiv EPOC+ is one of the most widely used wireless EEG headsets that has been employed in studies of emotion detection and self-reflection \cite{roo2016inner, ramirez2015musical, pham2012emotion}. It contains 14 EEG channels and covers all four brain lobes, which are the frontal lobe, parietal lobe, occipital lobe, and temporal lobe. The connection between EEG sensors and the scalp doesn't require liquid gel compared with traditional EEG caps, but the saline solution is needed. Thus, it can be categorized as a semi-dry sensor. The device is intended to be a versatile, generic EEG recording device that can be used for a variety of applications, and the EmotivPro SDK is compatible with both Windows and MAC systems. Emotiv EPOC+ has the highest spatial resolution among selected devices and it represents the most accurate and reliable consumer-grade EEG device.

MUSE 2 is an EEG headband that comes with a guided meditation application. The alpha band of the collected EEG data is utilized to infer users' meditation scores (the more the user is concentrated, the higher the meditation score is). It's been used by people who are interested in meditation or mindfulness training for its low cost and ease of use. It's also utilized for research purposes like emotion detection, stress detection, and brain-computer interface (BCI) control \cite{merrill2018scanning, bashivan2016mental, bhayee2016attentional, abujelala2016brain, jiang2019memento, bird2020cross, asif2019human, hohmann2020mynd}. The headband is capable of measuring reliable event-related potential (ERP) components \cite{krigolson2017choosing, krigolson2021using}, even though it has a low spatial resolution as the trade-off for better usability. But it was also criticized for the limitation of attention measurement in non-laboratory conditions  \cite{przegalinska2018muse}. There are 4 EEG channels based on the 10-20 system as shown in Table \ref{tab:devices}, which locate in between the temporal lobe, frontal lobe, and parietal lobe. Even though EEG sensors are dry electrodes, to make the setup time shorter and get a better connection, alcohol wipes are recommended.


mBrainTrain's Smartfones is a prototype of headphones with EEG sensors on side pads and the bridge \cite{kartali2019real}. It was developed based on the prototype of cEEGgrid \cite{bleichner2017concealed, debener2015unobtrusive}, which is validated in event-related potentials and neural oscillations. Unlike traditional EEG signals that are recorded from the scalp, Smartfones collect electrical signals from the area around the ears for highly portable and unobtrusive EEG data collection. It has 11 EEG channels: 3 electrodes are located in the central scalp according to the 10-20 system, while the remaining 8 electrodes are placed within the headphone pads and close to the temporal lobe. A saline solution is required for the connection between the sensors and the skin. The data streaming platform named Smarting Streamer is only compatible with Windows and a third-party Bluetooth interface is required. 

Neurable is a startup company focusing on developing headphones that are capable of collecting EEG data from EEG electrodes on the inner surface of the pads, through which it creates the everyday BCI platform. The prototype name is Enten \cite{alcaide2021eeg}. Experiments of the device show that it can detect auditory steady-state response (ASSR) and P300 ERP \cite{alcaide2021eeg}. It has 20 EEG channels around the ears (10 on each side), sensors with numbers 1 to 10 located around the right ear, and sensors with numbers 11 to 20 located around the left ear. These are non-standard EEG locations with a trade-off for ease of daily usage \cite{pereira2018cross}. Alcohol wipes are required for the connection with the skin. The SDK is developed by the CGX company and is only compatible with Windows.

When using the Emotiv EPOC+ and MUSE 2 devices (they don't have built-in earphones), participants were required to bring their own earphones or earbuds for music listening. Smartfones and Neurable devices are headphones ready to use, and the wire or Bluetooth that streams music is separate from the EEG data streaming interface.


\begin{figure*}[tbp!]
  \centering
  \includegraphics[width=0.77\linewidth]{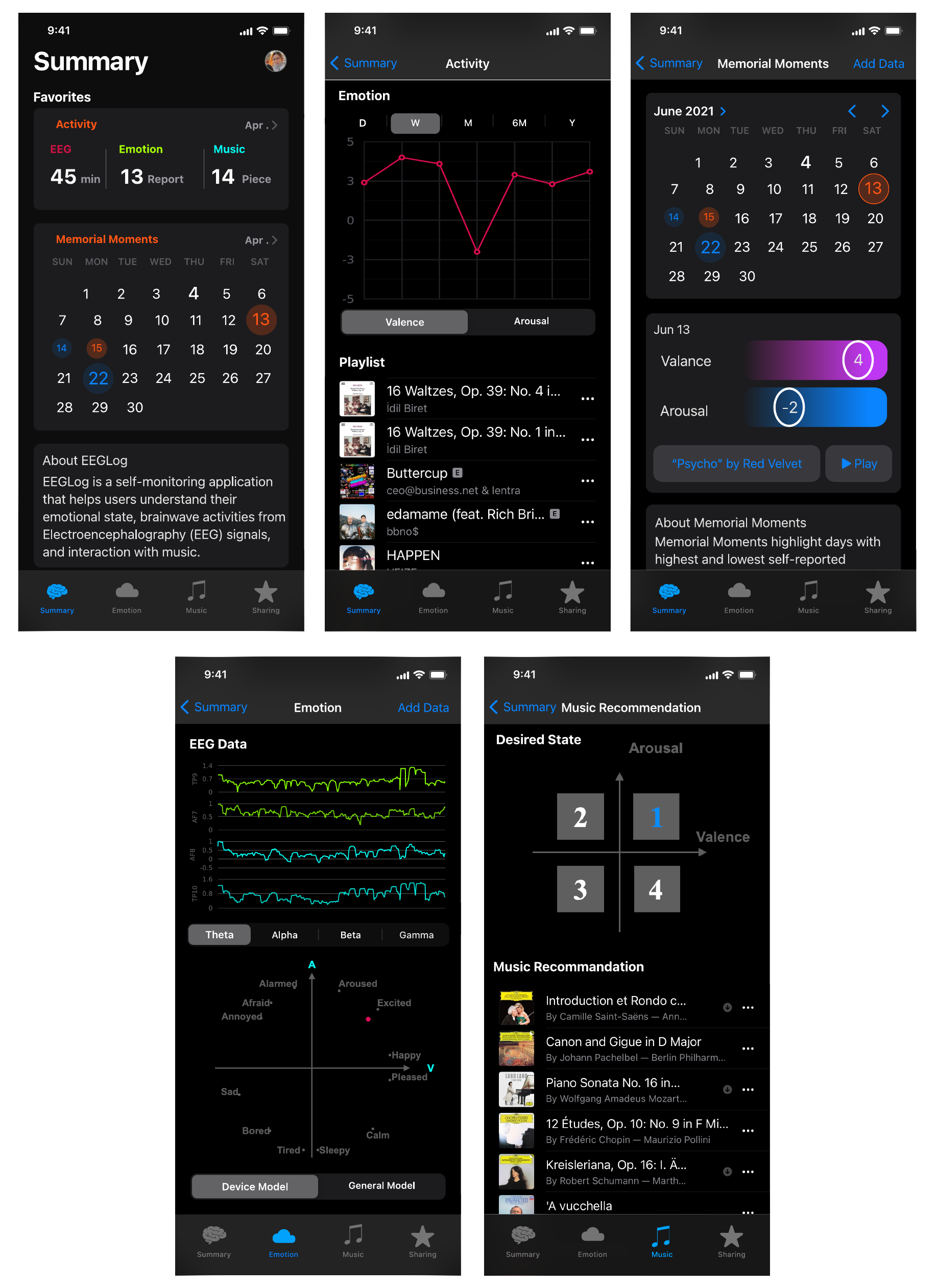}
  \caption{Exemplary windows of EEGLog. The upper-left page is the Summary window and users can check their Activity and Memorial Moments via upper-middle and upper-right windows. The lower pages are the Emotion Detection module and the Music Recommendation module.}
  \label{fig:log}
\end{figure*}


\subsection{EEGLog Application}
Self-monitoring applications have gained increasing attention recently because of people's growing interest in monitoring and quantifying both physical and mental health, like emotions. We provide three modules in EEGLog for users to reflect their emotional states on various days, and to interact with their preferred music based on their current emotional state. Users can review their past data in the module of ``Summary'' in the EEGLog as shown in Figure \ref{fig:log}, which includes activities of recorded EEG data, music pieces he/she listened to, emotion states that the user reported, as well as ``Memorial Moments'' concluded in a month. Also, two modules based on emotion learning algorithms named ``Emotion'' and ``Music'' are provided for users, which target emotion detection and music recommendation. Traditionally, Tracking users' emotions is based on users' manual inputting or related content like social posts and conversations. In this study, we propose to track the emotion based on the experience of listening to music. Such evoked emotion states are easier and less disturbing than unconditional self-report. Moreover, there is a ``Sharing'' module that we expect to achieve in the future, so that the user can share the emotion, memorial moments, recommended music playlists to their friends, and enhance the affective communication. Figure \ref{fig:log} shows the prototype of the EEGLog that is designed on Adobe Xd and based on the Apple Health structure.


\subsubsection{Summary}
Like ``Move'', ``Stands'', ``Steps'' in Apple's Health application, users can have a quick reference of their EEG data logging activity and highlights in the ``Summary'' module of EEGLog as shown in the first user-interface of Figure \ref{fig:log}. The first section of ``Summary'' is the ``Activity'', which shows the length of EEG data collected, the number of emotion reports that the user reported after listening to songs, and the number of songs that they listened to while logging the EEG data in a week. The user can view the reported Valence or Arousal score in different time spans when go to the `Activity'' interface with corresponding songs listed below as shown in the [] of Figure \ref{fig:log}. And the second section is the ``Memorial Moments'' which highlights days that the user reported the lowest and highest emotion scores (valence and arousal) in a month as shown in the [] of Figure \ref{fig:log}. The reported peak v/a values and the corresponding song are shown below when clicking on the day. It is an offline application used to reflect past data after data was collected.  

\subsubsection{Emotion Detection}
In the ``Emotion'' module, users can select an EEG segment or connect with the EEG device to detect the emotional state in real time. At the present stage, we have difficulties streaming EEG data from devices to any third-party platform because of compatibility issues, thus the real-time emotion detection is not available on EEGLog but is achieved with Python. Recorded EEG data is shown in frequency bands: theta ($\theta$: 4-7 Hz), alpha ($\alpha$: 8-13 Hz), beta ($\beta$: 14-30 Hz), and gamma ($\gamma$: 31-50 Hz). And with the selection of each band, the EEG signal is shown in the interface, and the detected emotion is marked below in Russell's circumplex model as shown in the [] of Figure \ref{fig:log}. Emotions are labeled into four classes based on positive and negative v/a values. The discreet emotional states shown in the model are only for reference. 

The emotion detection model is trained for each device based on the data collected from it. The reason is that the amount of self-collected data is relatively small and unbalanced for each participant. For example, participant 12 has 61 trial data, and 60 of them are reported with positive valence and positive arousal. Thus, we combine the data of different users using the same device to train the detection model. To distinguish with the general model trained with the public dataset DEAP \cite{koelstra2011deap}, we annotate it as the ``device model.'' At the preliminary stage, the data size is small for a generalized model. DEAP dataset \cite{koelstra2011deap} is a multi-modal dataset developed for emotion analysis, and we utilize it to present a more general result. DEAP collected data from 32 participants as they watched music videos for 40 trials, and participants reported their arousal and valence scores afterward. There are 32 channels of EEG data, and is downsampled to 128 Hz. Trials from the DEAP dataset are combined with self-collected data of each device to train the ``general model'' for each device. 

\begin{figure*}[h]
  \centering
  \includegraphics[width=0.55\linewidth]{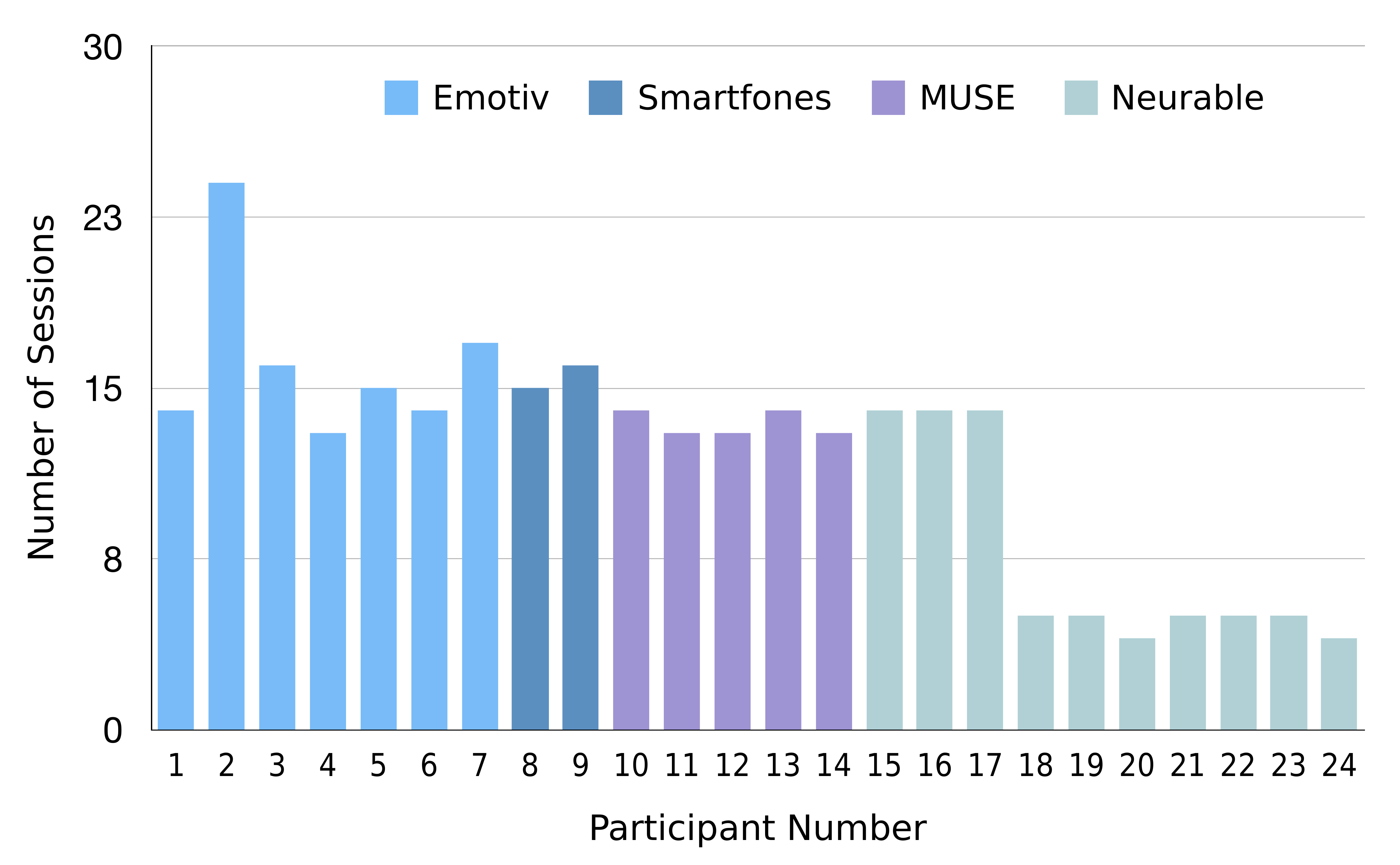}
  \caption{The number of sessions that participants 1 to 24 took part in of each device.}
  \label{fig:device}
\end{figure*}

\subsubsection{Music Recommendation}
Listening to music has been studied with self-regulatory goals for people, like changing, maintaining, or reinforcing emotions \cite{van2015listening, chen2007temporal, knobloch2002mood}. Music selection strategies are highly related to users' emotions. In the module of ``Music'', users can select a desired emotional state from four classes and get a recommended music playlist as shown in the [] of Figure \ref{fig:log}. The recommendation strategy is based on users' past data, which finds music pieces that have the same class of emotions as their desired one. Some studies proposed an online music recommendation system that aims to enhance the current emotional state based on the real-time EEG data or predefined a target emotional state for music recommendation \cite{ramirez2018eeg}. However, the emotional regulation experience is personal for each individual \cite{van2013exploring}, and we propose that users can explore and interact with their music-evoked emotions with choices of desired emotional states, no matter whether it is about strengthening, maintaining, or changing emotions. 


\section{Study}

\subsection{Participants}
The same experimental procedure was conducted for each individual. Participants 1 to 24 took part in the experiment of Emotiv EPOC+, MUSE 2, Smartfones, and Neurable, respectively, as shown in Figure \ref{fig:device}. The ages of participants are from $19-29$. And half of them are females. Among all participants, 16 participants accomplished 13 to 15 sessions of experiments, one participant accomplished 24 sessions, and the remaining 7 participants accomplished 4 to 5 sessions. All participants were recruited under the approval of the Internal Review Board (IRB) at [University is hidden for double-blinded review]. There is no clinical (including psychological and mental) problem reported by participants. 


\subsection{Procedure}
\begin{figure*}[tbp!]
  \centering
  \includegraphics[width=0.7\linewidth]{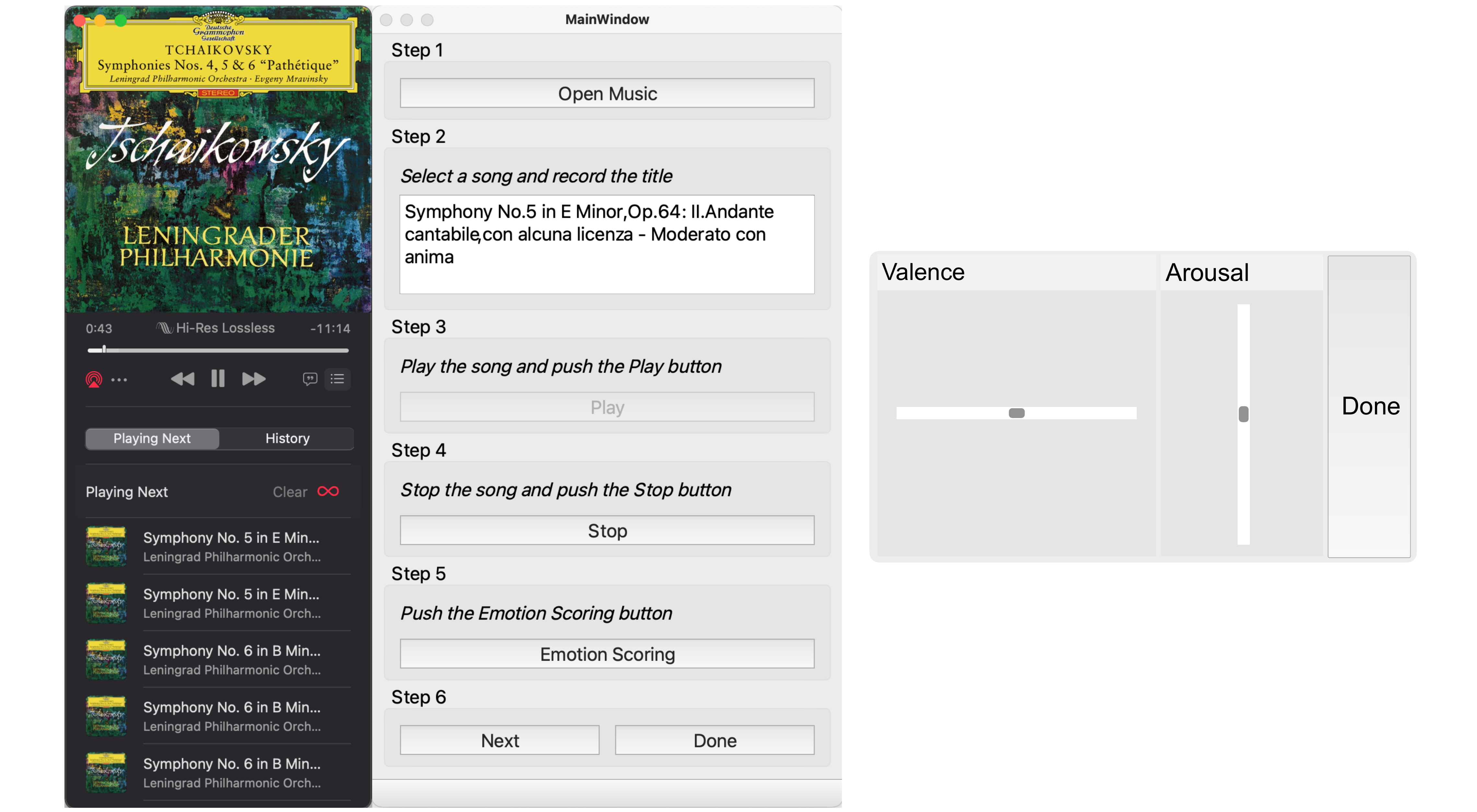}
  \caption{User-interfaces designed to save users' selections of music, and self-report of emotional states after listening to music.}
  \label{fig:expr}
\end{figure*}

Practically, a session is defined as a period that the user starts to listen to music until he/she stops listening. In our experiment, participants are instructed to use the EEG device and listen to music for at least four sessions on different days. The time of each session varied based on the number of songs they listened to, which varied from 3 to 12. In a session, each trial composes the steady state before listening to music, while listening to a music piece, and self-reporting after listening to the music piece. EEG data collection lasts for the whole experimental session. Song information and self-reports were collected by user interfaces designed for the experiment, as shown in Figure \ref{fig:expr}, which was driven by Python and designed with Qt. In the figure, the first interface connects with the music streaming platform for users to select songs, the name is recorded automatically. The interface on the right shows up when users click ``Emotion Scoring'' after listening to music, and the value of valence and arousal is recorded. Timestamps are saved when the user clicks on the ``Play'' button and the ``Stop'' button, which are used to segment EEG data.


The self-reporting of emotional states is based on Russell's two-dimensional (arousal and valence) emotion model \cite{russell1980circumplex}. Before the experiment, participants were asked to understand the two-dimensional model with its labeled discrete emotions, and with the help of the Self-Assessment Manikin (SAM) \cite{bradley1994measuring}. The valence bar and the arousal bar in the interface range from $-5$ to $5$, and users move the bar from the start point $0$. During the experiment, EEG data is streamed and saved by the device's SDK independently. 


\subsection{EEGLog Trial}
The offline emotion activity and emotion detection are available for all participants. And they reviewed their Memorial Moments and recommended music playlists by choosing the desired emotion after finishing experimental sessions. Only participants from the Emotiv EPOC+ group tried and evaluated the online Emotion Detection based on real-time EEG, because only EmotivPro SDK provides user-friendly real-time data streaming that works with Python.

\subsection{Emotion Detection}
We train and test the emotion detection model for each device individually. Directions of valence and arousal emotions are learned by binary classification models independently. We don't consider the regression model to learn the valence and arousal scores because people tend to report and interpret their emotions by how intense he/she feels instead of exact values. Besides, the regression model raises the emotion detection problem to another level that involves the boundary of emotions, which has never been precisely defined. The most reliable models either classify discrete emotions into classes of basic emotions or distinguish valence and arousal as binary variables. The Support Vector Machine with Radial Basis Function kernel (SVM-RBF) was employed in this study, due to its superior generalization capability and suitability for a smaller amount of training samples. Its effectiveness has also been well demonstrated in prior EEG-based emotion classification studies \cite{bazgir2018emotion}. We trained two models for each device, one with self-collected data of the device, and another one with a combination of self-collected data and a publicly available EEG dataset.

To expand the data size and show a more general result of detected emotions to users, EEG data from the DEAP dataset is added to train the general model. The problem is that we can only combine the data from electrodes at the same location or very close location as our EEG devices. As shown in Table \ref{tab:chan}, for the MUSE 2, the DEAP dataset doesn't have the same EEG channels as MUSE 2, so we choose the closest electrodes to represent those four channels, which are shown in parenthesis correspondingly in Table \ref{tab:chan}. For the Emotiv EPOC+ device, excluding electrodes located on the occipital lobe, there are 10 EEG channels that overlap with the DEAP dataset, all of which are chosen and used in our comparative analysis. The Smartfones headset has three overlapped electrodes (C3, C4, Cz), but the other eight electrodes around two ears are unique. Therefore we choose the closest electrodes T7, and T9, from DEAP to represent electrodes L2 and L3 in Smartfones. Electrodes of the Neurable Enten don't overlap with any electrodes of the DEAP device, so T7 and T9 are chosen to represent sensors 16 and 5. In conclusion, there are 4, 10, 5, and 2 channels of data for the general model trained on each device.


\begin{table}[]
\centering
\caption{Selected EEG channels of four wearable devices for general model training, and their corresponding channels in DEAP dataset. Electrodes in parenthesis are the closest substitutions in DEAP dataset.}
\label{tab:chan}
\begin{tabular}{ll}
\textbf{} & \multicolumn{1}{c}{\textbf{EEG channels}} \\ \hline
\textbf{MUSE 2} & TP9 (T7), AF7 (F7),  AF8 (F8), TP10 (T8) \\ \hline
\textbf{Emotiv EPOC+} & AF3, F7, F3, T7, P7, P8, T8, F4, F8, AF4 \\ \hline
\textbf{Smartfones} & L2 (T7), R2 (T8), C3, C4, Cz \\ \hline
\textbf{Neurable} & 16(T7), 5(T8) \\ \hline
\end{tabular}
\end{table}




\subsection{EEG Data Processing \& Feature Selection}
For the model training and testing, all the EEG data were high-pass filtered and low-pass filtered with thresholds of 2 Hz and 50 Hz. The power spectrum density (PSD) of the filtered data was calculated for the epoch of listening to music. It was divided into four bands: theta ($\theta$: 4-7 Hz), alpha ($\alpha$: 8-13 Hz), beta ($\beta$: 14-30 Hz), and gamma ($\gamma$: 31-50 Hz). The Delta band was not selected for the low-frequency components, including noises like eye blinks, head movement, and so on. Besides, the Delta wave is mainly studied for sleeping stages, consciousness, and cognitive performance, which is less related to emotions. Four devices are processed based on the sampling rate we configured as shown in Table \ref{tab:devices} individually, and feature vectors are concatenated by the PSD of each band and each channel. The resulting EEG feature vector has the dimension of $(4 \times number\ of\ channels)$. Electrodes located on the occipital lobe are not considered in this study because only the Emotiv EPOC+ device covers the lobe, and it's less relevant to emotion and acoustic stimulation. We use the same procedure to process EEG data in the DEAP dataset.

The dimension of the feature vector is different for each device, as they have different numbers of channels. For the device model, each device has more than 11 channels, which contains 44 features in total, except MUSE 2. And they should be decreased for the limited number of trials/samples. To select features, the greedy search algorithm Sequential Backward Selector (SBS) was utilized, which removes the feature one by one to maximize the performance. We set the number of features to 20 to satisfy the device with the least number of trials, and features for valence and arousal are selected individually. For the general model, selected channels are decreased are shown in Table \ref{tab:chan}. Only Emotiv EPOC+ has more than 20 features and is implemented feature selection by SBS.

\section{Results}

\begin{table*}[tbp!]
\centering
\caption{Traning and testing results of the device model and the general model.}
\begin{tabular}{llllllllll}
\toprule
 & \textbf{} & \multicolumn{4}{c}{\textbf{Device Model}} & \multicolumn{4}{c}{\textbf{General Model}} \\ \cline{3-10} 
\textbf{Devices} & \textbf{\#trails} & \textbf{v\_train} & \textbf{v\_test} & \textbf{a\_train} & \multicolumn{1}{l|}{\textbf{a\_test}} & \textbf{D\_v\_train} & \textbf{v\_test} & \textbf{D\_a\_train} & \textbf{a\_test} \\ \hline
\textbf{Muse 2} & 314 & \textbf{0.8326} & \textbf{0.8412} & 0.7610 & \multicolumn{1}{l|}{0.7143} & 0.9490 & 0.6032 & 0.9058 & 0.6944 \\ \hline
\textbf{Emotiv EPOC+} & 1284 & 0.6796 & 0.7198 & 0.6650 & \multicolumn{1}{l|}{0.6615} & 0.7918 & 0.5447 & 0.7532 & 0.5291 \\ \hline
\textbf{Smartfones} & 62 & 0.7755 & 0.7692 & 0.7346 & \multicolumn{1}{l|}{0.6923} & 0.9786 & 0.5400 & 0.9795 & 0.5385 \\ \hline
\textbf{Neurable} & 237 & 0.7903 & 0.7446 & \textbf{0.7634} & \multicolumn{1}{l|}{\textbf{0.7234}} & 0.7702 & 0.5561 & 0.7868 & 0.5936 \\ \bottomrule
\end{tabular}
\label{tab:test}
\end{table*}

\subsection{Emotion Detection}
Training and testing results of emotion detection models are shown in Table \ref{tab:test}. Accuracy of valence and arousal models trained by the EEG data collected from the same device is shown under \textit{v\_train}, \textit{v\_test}, \textit{a\_train}, and \textit{a\_test}. The number of samples used for each device model is shown under \textit{\#trials}. And The ratio for training and testing samples is $0.8/0.2$. The accuracy of the general model trained with the DEAP dataset is shown under \textit{D\_v\_train}, \textit{v\_test}, \textit{D\_a\_train}, and \textit{a\_test}. EEG data from the DEAP dataset is used for training only, testing is done with self-collected data. The number of trials for the general model of each device is \textit{\#trials} plus 1280, which is the number of the DEAP trials. 

Device models perform better overall for emotion prediction compared with the general models. The high training score of general models is because of the good performance of the DEAP dataset, which accounts for the major part of the training data. And the accuracy decreases when the self-collected data size increases, as shown in the result of the Emotiv EPOC+ device, it has the largest amount of trials and lowest \textit{D\_v\_train} and \textit{D\_a\_train}. The average testing results of four device models are $0.7688 \pm 0.0454$ for valence detection, and $0.6979 \pm 0.0239$ for arousal detection. The detection relies on EEG data and how accurately users annotate their emotions. Possible reasons that valence emotion detection is more accurate is that frequency bands of EEG data can distinguish valence emotion better than arousal emotion, and users are more certain about their valence emotion, compared with arousal emotion. 

Testing results of general models are less satisfactory overall. The accuracy is generally over $0.5$ and thus, the reference value of the general model is low for EEG-based emotion detection tasks. However, the training accuracy is high, which means that the video-stimulated emotions can be detected accurately by the EEG device that the DEAP uses. The low testing accuracy reveals a problem in that people's brain reactions would be different for the same emotion that was stimulated by different media. And compared with the music piece that is a one-dimensional signal, video clips contain more explicit components related to emotions. And it would be easier to evaluate the emotion when involved in the story and scene. The general model should be further explored by taking into account modalities, experiment designs, and so on.

\subsection{Device Comparison}
Among four devices, MUSE 2 has the highest training and testing results for valence emotion classification, and Neurable has the highest training and testing results for arousal emotion classification. And they are the two most comfortable and easy-to-set-up devices of all. Both demonstrate that non-standard locations of EEG electrodes record emotion-related brain activity and are good for daily emotion classification. EEG lifelogging based on consumer-grade EEG devices is feasible and reliable. Emotiv EPOC+ has the most standard electrodes and the number of trials, which is assumed to be the most accurate EEG device among all. However, the emotion detection performance is after the other three devices. The conclusion from this is that after users logged in more trials of data, it would be better to train their own personalized model instead of the device model. Users' emotional states and EEG data vary over time and it can mislead the model for other users. 

There are two major complaints from participants about how uncomfortable they wear the EEG device after a certain period. The first one is the tightness, for better connections between the scalp/skin and the electrodes, EEG devices are worn tight on the head, especially over the mastoid bone. And electrodes like Emotiv EPOC+ have a small contact surface that increases the pressure. Except for Neurable, all other devices were complained regarding the tightness, and the most complained one is Smartfones, which is tight and heavy, then Emotiv EPOC+, and the last one is MUSE 2. The second major complaint is the muggy or sticky feeling of the sensors. To connect with the scalp and skin, sensors like Smartfones and Emotiv EPOC+ are designed to absorb much of the saline solution. The saline solution drops through the face when setting up the device, and soaks hairs after usage. Material of Neurable's sensors also absorbs alcohol from wipes but it's less than the former two. MUSE 2 doesn't require any supplement but the connection process is harder and longer without alcohol wipes. In conclusion, all four devices have the wetness problem, and the most complained one is Smartfones, then Emotiv EPOC+ and Neurable, and the least complained one is MUSE 2. It shows that the device that can be worn for the longest time is Neurable, and the one that can be worn for the shortest time is Smartfones. The recorded acceptable time length of wearing these devices is shown in Table \ref{tab:devices}, which is a weakness when considering substituting headphones that can be worn much longer.



\subsection{Experiment Procedure \& Emotion Evaluation}
Experiment setting and procedure always impact participants' feelings while collecting data, including experimenters, unfamiliar devices, and the lab environment. They can distract users and change their emotions. All experiments were done in the lab with one experimenter. After the experiment, we interviewed participants with questions of what are their feelings about the EEG data login procedure, and how's their experience with the emotion evaluation. Participant 14 said that ``\textit{Sometimes I was a little irritated as I know that the experimenter is waiting for me when I don't know what my valence or arousal emotion score should be.}'' And Participant 8 said ``\textit{For the first few days, I was distracted a bit by the presence of the experimenter, but when I got familiar with the whole process, I was able to focus}''. From the feedback of users, devices don't impact participants' emotions, but people around them give them pressure. And this problem can be solved when users use the EEG device at home by themselves.

The difficulty extent of emotion evaluation differs among different people. It's interesting to find that among participants of this study, users who have very stable and less intense emotions feel more confident about where their emotions are, which conflicts with common sense that it's easier to measure the emotion when it is more intense. Self-report of participant 12 is one of the stablest, which is $0.6639 \pm 0.6027$ and $0.5311 \pm 0.4135$ for valence and arousal through all his trials. He/she delivers that ``\textit{I don't have any problems when I measure the emotional states, no. I can usually gauge how I feel at most if not all moments in time.}''. And another participant whose emotion is stable takes the least time to report valence and arousal emotions through trials. Most participants mentioned that they don't know what the exact feeling is but can track where it is after a few sessions. Participant 13 says ``\textit{I would say that I was marking what I think I was feeling, I still don’t know if that was what I actually felt. But that's what I did for all sessions}''. The practice of measuring emotions with the valence-arousal model helps users conceptualize their emotions to some extent. Participant 11 said that ``\textit{I think learning about the continuous model has definitely changed how I see my own emotions, and having to focus on them in real-time and report them has sort of made me do it more often in my own life, like stop and kind of analyze how I am feeling in terms like that}''. And participant 3 said that ``\textit{it becomes a habit and makes me equipped when communicating how I feel to others}''.

\subsection{Music Agent \& Emotion Regulation}
Music plays different roles and permeates every aspect of human life. All participants expressed that they like to listen to music in various situations, making the experiment easy and enjoyable. Participant 5 said that ``\textit{I am almost constantly listening to music when I can unless I am doing something where I can’t}''. And Participant 1 also listens to music all the time he/she works out, does work, or relaxes. Some people only listen to music when they are in a certain mood or situation. Participant 10 said that ``\textit{ I would say that the only time I listen to music is before I start on particularly hard work, a prayer song can help out. And it can actually really irritate me when I need to pay attention to driving or working}''. And participant 12 said that ``\textit{music listening is more of a selected task, so I wouldn't put music on if I am working on specific tasks. But if I’m like I’m in a good mood and I feel like I could get into a good groove, I will put on like an album or something}''. For participants who only listen to music in specific situations, EEGLog would be biased and only contains specific emotional states and songs. And it is against the concept of all-encompassing capture of lifelogging.

Music provides a way for emotional expression and communication. Mood enhancement is one of the well-documented self-regulatory goals. People are not always motivated to seek positive and joyful music \cite{van2013exploring}. Music pieces chosen to amplify their emotions are indicating of how they feel. Participant 11 said that ``\textit{sometimes I choose a song that amplifies the feeling instead of changing it, so if I am already kind of feeling sad, then I listen to music that can make me feel it more strongly. I don’t necessarily listen to music to change how I feel but more to reflect how I feel}.'' Generally, participants' selection of music depends on how they feel at that time, which should be explored for music recommendation strategies. Participant 17 said that ``\textit{I believe there are many emotional components are non-verbal, and I'm willing to share my music and friends can know more of my personalities and preferences, also I'm not an extravert}''. 

Familiar music pieces can provoke memory that helps self-reflection with remembering. Participants reviewed their memorial moments on EEGLog before the interview. Participant 2 said that ``\textit{I listen to soundtracks that I like from dramas I watched, the one that corresponds with the highest valence is one of my favorites, and I can recall the story and main characters once I listened to it}''. Participant 3 said ``\textit{the song with the highest valence is my favorite Christmas song, and the joyful atmosphere suddenly comes to my memory and I know that it's exactly can be the song with the highest valence}''. Not just positive and pleasant feelings can be recalled and reflected by participants. The song that comes with sad or unpleasant feelings is also reviewed and participants deliver they have different feelings compared with before. ``\textit{I'm surprised this song has the lowest valence and low arousal, I recall it makes me calm but also a bit pleasant for other days}''.

\subsection{EEGLog Experience}
One of the main features of EEGLog is the EEG analysis, which reveals users' emotions for a song. Participants' feedback shows that the memorial moments are mostly in alignment with their thoughts as one said ``\textit{My go-to for a mood lifter ended up showing the highest valence and the song that excites me the most correlated to the highest arousal}'', but it's also confusing when the information contradicts with their memory and interpretation as ``\textit{the surprising result is the song that produced the highest alpha power is a song I enjoy and listen to a lot when I’m stressed, I would never have guessed that would produce the highest alpha power, if it relates to a more intense feeling}''. Trends can be found by users when they are reviewing their data. Participant 12 mentioned that ``\textit{I noticed that my arousal and valence scores were higher for the songs that I was more familiar with versus lower for both songs that I was unfamiliar with}''. It is expected that trends can be concluded by users when they use EEGLog longer, and the application can summarize trends with more sufficient data. Like analyzing emotions towards the same song that was listened in different days. Participants also try to understand the alpha power of brainwaves with new discoveries from the application. Participant 5 said that ``\textit{I'm not sure how alpha wave relates to my emotion, but my least favorite song ended up correlating to the lowest alpha power and the lowest valence}''. Participant 1 said that ``\textit{I know that alpha wave is kind of related with the resting state, and the highest of it corresponds to one of my favorite songs, which makes me more assured that how inherently I like it hh}.'' 

Overall, participants provided positive feedback on EEGLog as one mentioned ``\textit{It is helpful and gives a good overview as to what I need to listen to, it was amazing to collect data and have a program that tells you what is best when I have no idea what works}'', and another participant said ``\textit{I had no surprises about initial emotions for each day, but it feels good that something records my emotions}''. Regarding music recommendations of streaming platforms, some users entirely rely on them and some have doubts about them. And it is the same for our recommendation application, for the taste is very elusive and varied. ``\textit{I would say some songs are exactly what I want to listen to at this time, but some are not thought about, it's very interesting that there is a platform recommending music based on my brainwave}.''

Participants 1, 2, 3, 6, and 7 took part in online trials of Emotion Detection and Music Recommendation. We assume users use these two applications on specific occasions, like when they are curious about the current emotion, thus numerous trials are unnecessary. After seeing results from Emotion Detection, participant 1 mentioned that ``\textit{In general it detects my emotion but it distracts me when different results show by two models, and it makes me more unsure about what exact emotion I have}''. The same question was raised by another participant. We think that referring to the predicted emotions of models to criticize their own interpretation in real-time is not an entry-level usage. Therefore, users may think it is useless at the beginning, but the real-time objective analysis has the potential for users' self-regulation because the feeling shifts and cannot be reproduced. We assume after practice and getting used to the conceptualized feeling, users will be more confident about their interpretations and judgment of their emotional states.




\section{Discussion}

\subsection{Interaction with Music and EEG Data}
People's EEG dynamically changes with the music. However, the self-report of emotional state only represents one state after listening to music. Sometimes, users may enjoy the beginning of the music but start to dislike the remaining part and vice versa. Thus, EEG features extracted from the epoch when they are listening to music are not perfect for learning afterwards emotion. The interaction between users and music is changing too. Users' comprehension of their emotions and music tastes may be different over time. With more data being collected, the system should be able to conclude users' gradual and subtle changes with more crafted functions and algorithms. Besides, music pieces would influence the user's brainwave in terms of entrainment besides emotional change, which has an impact on emotion detection. 

The music lyrics, melodies, rhythms, components, etc., are not analyzed in this study. However, genres or high-level features of music pieces can be informative towards participants corresponding emotional states, and they can be useful for users to review their music selection behaviors through days plus recommendation strategies. Moreover, familiarity with the music is a variable that influences participants' brain activity and emotions. It is unsure whether the new song should be separated from familiar songs, especially for Memorial Moments, because the emotion raised by freshness will diminish when users recall it. And when recommending music based on similar brain activities or desired emotional states, new songs would no longer stimulate the same emotions. 

Evoking emotions by music pieces is different from emotion regulation, which aims to gear towards the specific goal of maintaining a comfortable emotional state \cite{blaustein2018treating}. The emotion regulation involved in this study is based on users' own choice of emotional states they want. However, we didn't track their emotion after listening to the recommended songs besides users' remarks on the playlist in general. So we are not able to evaluate the emotion regulation impact of EEGLog. Besides, users' emotional response toward a song depends on how they feel right before listening to music. Thus, referring to past self-report to recommend music would be outdated and improper. 

EEG signals are reliable for emotion detection and able to reveal neural activities like cognitive state, focus state, stress level, etc. However, EEG data can be distorted easily by movements. Thus users are suggested to use the sensor when they are in a steady state. Scenarios like users singing the song with the music, following beats, shaking heads, or exercising and other activities would raise problems for EEG data processing and the risk of rejecting those data segments. And it's still the weakness of the EEG-based system when introducing it to daily life.  
 
\subsection{Application Customization}
On one aspect, EEGLog presents participants' personal data, but it doesn't have functionalities that can be customized by users. The application we provide is basic and simple. To further help users with self-reflection and emotion regulation, more inputs and functions should be included for users' options. For example, some people would be interested in the emotion change rule of a day, then data collected at a different time of the day should be analyzed and summarized, like morning, afternoon, evening, etc. It would also be helpful for people can log in with information like physical conditions, caffeine intake, and so on. Then the EEG analysis will not only be based on the self-report of valence and arousal emotions. For female users, ``cycle tracking'' can be a critical index for emotion changes for premenstrual syndrome (PMS) is a wide problem, and lifelogging EEG data has the potential to reveal patterns of cycles. Other health factors like heart rate, and physical activities would relate to users' emotions and brain activities, and thus can contribute to more considerable analyses. And it would be a big advantage if the data in the Health application and EEGLog can be shared mutually.

The emotion detection model is based on all the data from the same device at this stage, and we see the decreased accuracy when the data size increases. With more data logged in, the unbalance problem will still exist if users only select music that makes them feel a specific emotion. The emotion where the valence is negative and arousal is positive rarely happened compared to the other three. The solution would still be based on the public dataset that included EEG data corresponding to all classes of emotions or explore a more suitable emotion model instead of a binary classification of valence and arousal. Plus, the explosion of data is a challenge with more extended usage. However, compared with photos or videos, EEG data is less vulnerable to the storage shortage issue. 


\section{Limitation and Future Work}
One limitation of the present study was the length of time participants used EEG devices and EEGLog. The time span limits the data shown in Activity and Memorial Moments so that users can only review recent emotional states and music histories. It is possible that the `memorial moments' during a short period of time are not memorials for users. With long-term usage, users may have different experiences of recalling and reviewing past data, and the function can pick up more meaningful moments. Music recommended for participants is also from their latest history. Even though there are songs they listen to repetitively, the recommendation strategies can be improved with a longer music listening history.

The application doesn't take users' feedback and update results at this stage. However, users should be able to interact with EEGLog in time in the future, like reporting their emotions after reviewing the results of Emotion Detection. Or input opinions about the selected memorial moments and music pieces. A diary input can also be useful for future self-recollection. Besides, other frequency bands related to drowsiness, stress, and alertness can be provided to users for indicative information. And primitive functions like a short-time Fourier transform (STFT) can be employed to show EEG change over time within a song. Besides providing the song name, comparing the time-frequency features of music to EEG features is a further step that can be informative and interesting to users. And the specific music segment can be highlighted based on neural activities when the music piece is analyzed.

\section{Conclusion}
In this study, we propose EEGLog, the first system of its kind that introduces EEG logging when users listen to music. Given EEGLog's compatibility with any EEG device, and we test four COTS consumer-grade EEG devices with data collected from 24 participants. The logged data is analyzed offline and presented in four different applications of EEGLog for users' self-reflection and emotion regulation. Users took part in the EEG data log-in for 4 to 24 days and used EEGLog to review their emotions, EEG activities, and music history, after which they provided valuable remarks and perceptions of the whole experience. We contribute to the examination, introspection, and inspection of this novel idea and application. It is anticipated that this research can guide people to think and care more about their emotions and brain activities related to their daily life. The neural-centric interactions are still largely under-explored in the community, and we hope this study can encourage researchers to explore human self-tracking and reflection from a new perspective with the emerging EEG devices in the market. 

\section*{Acknowledgment}
This material is based upon work supported by the National Science Foundation under Grant No. CNS-1840790. Any opinions, findings, and conclusions or recommendations expressed in this material are those of the author(s) and do not necessarily reflect the views of the National Science Foundation.

\bibliographystyle{unsrtnat}
\bibliography{references}  






\end{document}